\documentstyle[twoside,fleqn,espcrc2,epsf]{article}

\newcommand{\be}{\begin{equation}}
\newcommand{\ee}{\end{equation}}
\newcommand{\R}{{\bf R}}

\newcommand{\AmS}{{\protect\the\textfont2
  A\kern-.1667em\lower.5ex\hbox{M}\kern-.125emS}}

\title{The Running Coupling from SU(3) Gauge Theory}

\author{G.S.~Bali and K.~Schilling\thanks{Work supported
        by EC project SC1*-CT91-0642.}\vskip\baselineskip
        Fachbereich Physik,
        Bergische Universit\"at, Gesamthochschule Wuppertal\\
        Gau\ss{}stra\ss{}e 20, 5600 Wuppertal, Germany}

\begin{document}

\begin{abstract}
We present high precision results on the static $q\bar q$ potential
on $32^4$ and smaller lattices, using
the standard Wilson action at $\beta = 6.0, 6.2, 6.4$,
and $6.8$ on the Connection Machine CM-2.
Within our statistical errors ($\approx 1\%$) we did not observe
any finite size effects affecting the potential values, on
varying the spatial lattice extent from 0.9 fm up to 3.3 fm.
We find violations of asymptotic scaling in the bare coupling
up to $\beta=6.8$. We demonstrate that scaling violations
on the string tension can be considerably reduced by
introducing effective coupling schemes, which allow for a safer
extrapolation of $\Lambda_L$ to its continuum value.
We are also able to see and to quantify the running of the coupling
from the interquark force. From this we extract the ratio
$\sqrt{\sigma}/\Lambda_L$.
Both methods yield consistent values for the $\Lambda$-parameter:
$\Lambda_{\overline{MS}}=0.558_{-0.007}^{+0.017}\times\sqrt{\sigma}
=246_{-3}^{+7}$MeV.
\end{abstract}

% typeset front matter (including abstract)
\maketitle

\section{Introduction}
It is of great interest to extend lattice techniques into a regime
where perturbative methods can be applied~\cite{chris,UKQCD,mack,su2schroed}.
We achieve a lattice spacing as small as
$a\approx (6$ GeV$)^{-1}$ on a $32^4$ lattice at $\beta = 6.8$.
{}From the static $SU(3)$ quark-antiquark potential we can extract
the running coupling ``constant'' within an accuracy that competes
with current (real QCD) experiments.

\section{Simulation}
In our simulations~\cite{wir,wir2} we have realized lattice
volumes of $L_S^3\times L_T=32^4$ for the
$\beta$-values 6.0, 6.4, and 6.8. and a $24^3\times
32$ lattice for $\beta=6.2$. The corresponding
spatial lattice extents in order of increasing $\beta$ are:
3.25, 2.74, 1.74, and 1.05 fm with the physical scale set by
the value $\sqrt{\sigma}=440$ MeV for the
string tension.
In order to investigate finite size effects (FSE)
we simulated, moreover, a $16^4$ lattice at $\beta=6.0$, $16^3\times 32$,
$24^3\times 32$, and $32^3\times 16$ lattices at $\beta=6.4$, and a
$16^3\times 64$ lattice at $\beta = 6.8$.

We used the hybrid overrelaxation algorithm for updating the gauge fields,
as advocated in Kennedy's review talk~\cite{kennedy}.
Great care was taken in optimizing
the overrelaxation step~\cite{wir}.

To improve the projection of our operators onto the
$q\bar q$ ground state we applied an iterative local gauge covariant smoothing
procedure~\cite{wir2} on the spatial links before constructing
Wilson loops. Overlaps of $95(80)\%$ for small (large) spatial
$q\bar q$ separations $R$ were reached.
Various smoothened on- and off-axis Wilson loops were
measured every 100
sweeps. This separation was found to be sufficient to guarantee an
(almost) independent sampling. Remaining autocorrelation effects were
healed by blocking the data into bins of reasonable length before
statistical analysis. On all volumes ${\cal O}(100)$ measurements were
performed.

The spatial $q\bar q$ separations $\R = n{\bf e}_i$
with ${\bf e}_i = (1,0,0),(1,1,0),(2,1,0),(1,1,1),(2,1,1),(2,2,1)$
were realized.
$n$ was increased up to $L_S/2$ for $i = 1,2,4$, and up to
$L_S/4$ for the remaining directions. Altogether this yields 72
different values of $\R$ on the $32^3\times L_T$ lattices. The
time separations $T=1,2,\ldots,10$ were used to test exponential
behaviour and for extraction of the potential values.

\section{Results}
\begin{figure}[htb]
\epsfbox{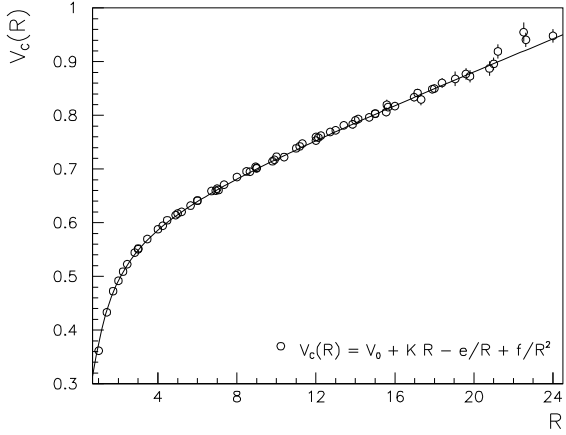}
\caption{Potential $V_C(R)$ at $\beta=6.4$, $V=32^4$.}
\end{figure}
\begin{table}[hbt]
\setlength{\tabcolsep}{1.5pc}
\newlength{\digitwidth} \settowidth{\digitwidth}{\rm 0}
\catcode`?=\active \def?{\kern\digitwidth}
\caption{Fit parameters (see Eq.~(1)).}
\label{fitres}
\begin{tabular}{lrr}
\hline
&$\beta=6.0$&$\beta=6.4$\\
\hline
$K$&0.0513(25)&0.0148(3)\\
$e$&0.275(28)&0.315(15)\\
$V_0$&0.636(10)&0.601(4)\\
$l$&0.64(12)&0.56(6)\\
$f$&0.041(58)&0.075(18)\\
\hline
&$\beta=6.5$~\cite{UKQCD}&$\beta=6.8$\\
\hline
$K$&0.0114(2)&0.0053(2)\\
$e$&0.311(14)&0.311(10)\\
$V_0$&---&0.549(2)\\
$l$&0.64(6)&0.56(4)\\
$f$&0.067(13)&0.094(13)\\
\hline
\end{tabular}
\end{table}

{\bf $q\bar q$ potential:}
We follow Chris Michael~\cite{chris} and start from the ansatz
\be
\label{eq-pot}
V(\R)=V_0+KR-e\left(\frac{1-l}{R} + l\,
G_L(\R)\right)+\frac{f}{R^2}.
\ee
The lattice propagator for the one gluon exchange $G_L(\R)$
has been calculated in the large volume limit.
The parameter $l$ controls violations of rotational
symmetry within this ansatz. The term $f/R^2$ simulates
deviations from a pure Coulomb behaviour and is expected to be positive
from asymptotic freedom, and to increase with the energy resolution.

In Figure~1 the lattice corrected data points $V_C(R)=V(\R)+el\left(
G_L(\R)-1/R\right)$ are plotted together with the interpolating fit curve.
The figure demonstrates the success of ansatz Eq.~(1) as all data
points line up on a smooth curve. Our potential fits yield $\chi^2/N_{DF}<1$
as long as the first two data points are
excluded.

We have included the corresponding
fit parameters from the recent UKQCD~\cite{UKQCD} investigation on a
$36^4$ lattice at $\beta=6.5$ in Table~\ref{fitres}.
For $\beta\geq 6.4$ all Coulomb coefficients $e$ are
definitely different (but remarkably stable) from the string
vibration value $\pi/12\approx 0.262$. $V_0$ decreases with the coupling.
We emphasize that the parameter $f$ is established to
increase with $\beta$ as expected. $l$ seems to vary only slightly.

{\bf Finite size effects:}
On varying the lattice extent at constant lattice spacing from 1.7 to 3.3 fm
($\beta=6.0$), and from 0.9 up to 1.8 fm ($\beta=6.4$) we found no
evidence of FSE on our potential data within statistical accuracy
($\approx 1\%$). However, a comparison of results on
the $\beta=6.8$ lattices (0.53 and 1.05 fm) shows deviations
(up to $3\sigma$), mainly for
small $R$ separations where statistical errors are small.
We conclude that a lattice extent $aL_S>0.9$ fm $\approx
2/\sqrt{\sigma}$ suffices in our case.
The same limit was also found in
Ref.~\cite{Perantonis}. This condition coincides with the
transition into the deconfined phase at
$aL_T>1/T_C\approx 2/\sqrt{\sigma}$.

Note, that a small volume can also cause problems of another kind:
It is difficult to fit the potential to a small number of
data points with some confidence. This effect becomes dominant for
$\beta\leq 5.9$ where the physical limit is $L_S \geq 8$.
In any case, it pays to work on a large lattice.
The computational task (except memory
requirement) remains comparable because self averaging suppresses
fluctuations and reduces the number of required Monte Carlo sweeps.
\begin{figure}[htb]
\epsfbox{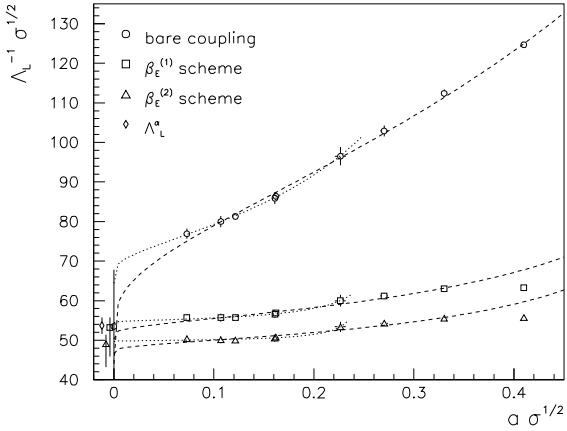}
\caption{Continuum extrapolations of $\Lambda_L^{-1}$.}
\end{figure}
\begin{table}[hbt]
\setlength{\tabcolsep}{0.7pc}
\settowidth{\digitwidth}{\rm 0}
\catcode`?=\active \def?{\kern\digitwidth}
\caption{$\Lambda_L^{-1}$ in units of the string tension.}
\label{tab-scal}
\begin{tabular}{lrrr}
\hline
$\beta$~~Ref&$\sqrt{\sigma}/\Lambda_L$&
$\sqrt{\sigma}/\Lambda_L^{(1)}$&$\sqrt{\sigma}/\Lambda_L^{(2)}$\\\hline
5.7\cite{MTC}&124.7 (07)&63.3 (04)&55.7 (03)\\
5.8\cite{MTC}&112.4 (10)&63.0 (06)&55.6 (05)\\
5.9\cite{MTC}&102.9 (14)&61.2 (08)&54.3 (07)\\
6.0& 96.5 (23)&60.0 (15)&53.4 (13)\\
6.2& 86.4 (10)&56.9 (07)&50.8 (06)\\
6.4& 81.3 (08)&55.7 (05)&50.0 (05)\\
6.5~\cite{UKQCD}& 80.0 (14)&55.7 (10)&50.1 (09)\\
6.8& 76.9 (13)&55.7 (09)&50.4 (08)\\\hline
$\infty$&$54^{+18}_{-15}$&$53.2^{+2.6}_{-7.3}$
&$49.1^{+2.3}_{-5.9}$\\\hline
\end{tabular}
\end{table}

{\bf Scaling:}
Within our range of lattice spacings, scaling of
physical quantities works nicely: In the present
investigation all potential data, measured at different $\beta$ values,
fall onto a single curve when scaled to $V_0=0$, and $K=1$.
However, though scaling seems to be restored, it still does not
follow the perturbatively expected two-loop behaviour
$a_{\sigma}=f(\beta)/\Lambda_L$ with
$a_{\sigma}=\sqrt{K}/\sqrt{\sigma}$, and $f(\beta)=$~the integrated
two-loop $\beta$-function. This is reflected in the first column of
Table~\ref{tab-scal}. It was argued that the lack of asymptotic
acaling may be caused by a bad
choice of expansion parameter. For remedy, a variety of
schemes with redefined expansion parameters have been proposed in the
past~\cite{parisi,mack2}.

Here, we attempt to achieve an improvement by using the (inverse)
coupling $\beta_E$ that has originally been invented by
Parisi~\cite{parisi}, and
successfully applied in Ref.~\cite{fing} to gauge theories. This
coupling is defined by truncating the weak order expansion of the
plaquette (measured in the Monte Carlo simulation) after the first order
term, and inverting the resulting relation.
As a check, we investigate an alternative
$\beta_E^{(2)}$ scheme~\cite{wir2},
defined by truncating the expansion after the second order term,
instead. The improvement can be seen from
Table~\ref{tab-scal} where $\Lambda_L^{(1)}$
corresponds to the $\beta_E$, and $\Lambda_L^{(2)}$ to the
$\beta_E^{(2)}$ scheme.

Extrapolations to the continuum limit
were done by fitting the data to the leading order expectation
$\Lambda_L^{-1}(a)=\Lambda_L^{-1}(0)+C/(\sqrt{\sigma}\ln(Da\sqrt{\sigma}))$.
There is a considerable bandwidth in the extrapolations, depending on
the number of points included in the fits.
The effective schemes help to decrease this uncertainty and teach us
that linear extrapolations are misleading. Our best guestimate for
$\Lambda_L(0)$ is $\sqrt{\sigma}=50.8^{+1.0}_{-4.6}\Lambda_L$.

An interesting program would be to take other nonperturbative quantities
(like small Wilson loops) that can be measured accurately on small
lattices, and use them to define different effective couplings.
Since one can translate, at a given $\beta$, one scheme
into another, self consistent extrapolations (and
interpolations) become possible.
By exploiting all available nonperturbative information, one can
in this way reconstruct the underlying $\beta$-function beyond the
two-loop approximation.

{\bf Running coupling:}
So far we have computed a mass ($\sqrt{\sigma}$) as function
of a coupling. In the following, we reverse this procedure and
determine the coupling $\alpha_{q\bar q}(R)$. This is
achieved~\cite{chris}
by numerically differentiating the (lattice corrected) potential,
and multiplying the resulting interquark force with $\frac{3}{4}R^2$.
The results (Figure~3) are in fairly good agreement with
$\alpha_{\overline{MS}}(a^{-1})$
($\alpha_{\overline{MS}}(q)=\alpha_{q\bar q}(r)+0.05\alpha_{q\bar
q}^2(r)+\cdots$ with $q=1/r$), calculated from the
scaling of the string tension in the manner proposed in Ref.~\cite{mack2}.
\begin{figure}[htb]
\epsfbox{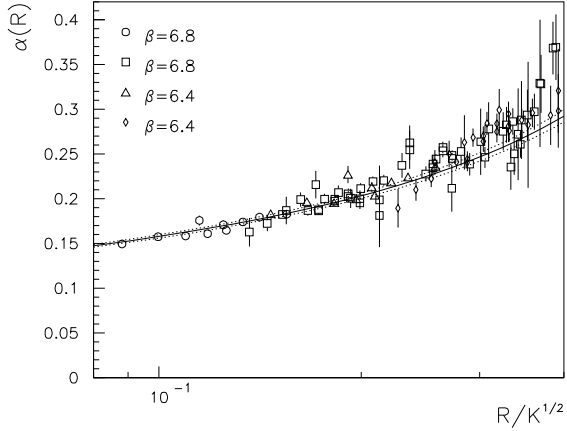}
\caption{The running coupling $\alpha_{q\bar q}(R)$.}
\end{figure}
\begin{figure}[htb]
\epsfbox{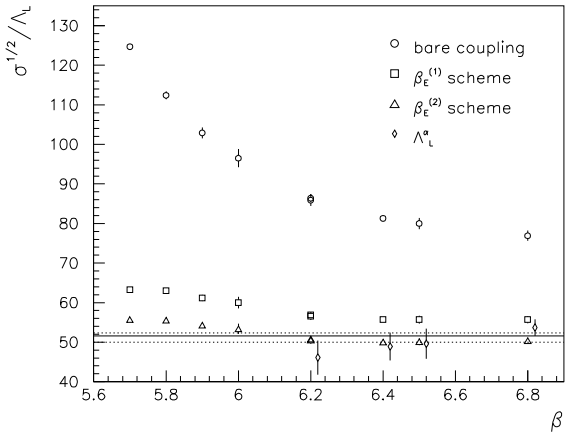}
\caption{Violations of asymptotic scaling.}
\end{figure}

We fit our data to the expectation
$\alpha_{q\bar q}^{-1}(R)
=4\pi\left(b_0\ln(Ra\Lambda_R)^{-2}+b_1/b_0\ln\ln(Ra\Lambda_R)^{-2}\right)$.
Exploiting the relation $\Lambda_R=30.19\Lambda_L$~\cite{billoire} we
arrive at $\sqrt{\sigma}=53.7(2.1)\Lambda_L^{\alpha}$ which is in agreement
with the value extrapolated from the scaling of the string tension.
We find the data to be in good agreement with the perturbative formula
for $q\geq 5\sqrt{\sigma}\approx 2$ GeV.
At small energies (large $R$) our data tends to lie above the
two-loop expression since $\alpha_{q\bar q}\propto 1/q^2 (q\rightarrow
0)$.

Averaging the two independently calculated $\Lambda$ parameters gives
$\sqrt{\sigma}=51.6^{+0.7}_{-1.6}\Lambda_L$ or
\be
\Lambda_{\overline{MS}}=0.558^{+17}_{-
7}\sqrt{\sigma}=246^{+7}_{-3}{\rm MeV.}
\ee
The approach towards this asymptotic limit for different definitions
of effective couplings is illustrated in Figure~4. We observe that the
values for $\Lambda_L^{-1}$, extracted from the
running of the coupling~\cite{chris,UKQCD,wir2},
tend to increase with the momentum cutoff.

\section{Conclusion}
For our results, it has been important to study both infrared,
and ultraviolett aspects in order to verify the reliability of the
continuum extrapolation. We might say that we have been lucky to get
hold of {\em asymptotia} within our means.
This is due to the discovery that the
running coupling ``constant'' is well described within this theory
by the two-loop formula down to a scale of 1--2~GeV.

If nature continues to be nice to us it is possible to predict
experimental numbers like $\alpha_S(M_Z)$ or
$\Lambda_{\overline{MS}}^{(4)}$
as explained in Ref.~\cite{mack}. In order to remove additional systematic
uncertainties caused by the fact that
experimentalists still have not discovered how
to switch off some quark flavours and deliver
us a quenched value for the string tension, it is preferable to repeat this
study in full QCD on the level of TERAFLOPS power. Meanwhile,
further improvements of lattice techniques are
of great interest.

\end{document}